# Wavelet analysis of corneal endothelial electrical potential difference reveals cyclic operation of the secretory mechanism

Cacace V. I.[1#], Montalbetti N.[1#], Kusnier C.[1], Gomez M. P.[2], Fischbarg J[1*].

1- Institute of Cardiological Investigations, University of Buenos Aires, and CONICET, Argentina.
2-National Commission of Atomic Energy, Group of Elastic Waves, and National Technological University, Buenos Aires, Argentina

# CV and MN contributed equally to this work

*Corresponding author:

Jorge Fischbarg, MD, PhD
Inst. of Cardiological Investigations "A.C. Taquini" (ININCA)-UBA-CONICET
Marcelo T. de Alvear 2270, C1122AAJ Buenos Aires, Argentina
Telephone: 011-5411-4508-3885
Email: fischbargj@fmed.uba.ar

Running title: wavelet analysis of corneal endothelial electrical potential difference

ABSTRACT

There is evidence that the electrical potential difference of corneal endothelium (TEPD) is related electro-osmotically to fluid transport. Hence, determination of the TEPD would serve as a measure of the fluid movement. The oscillatory nature of the TEPD has been recognized recently using the Fourier transform; the oscillations of the highest amplitude were linked to the operation of electrogenic sodium-bicarbonate cotransporters. However, no time localization of that activity could be obtained with the Fourier methodology utilized.

For that reason we now characterize the TEPD using wavelet analysis for the first time in the epithelial physiology field, with the aim to localize in time the variations in TEPD.

We find that the high-amplitude oscillations of the TEPD are cyclic, with a period of $4.6 \pm 0.4$ s in the average (n=4). The wavelet power value at the peak of such oscillations is $1.5 \pm 0.1$ $mV^2$ Hz in the average (n=4), and is remarkably narrow in its distribution.

## INTRODUCTION

There is evidence that the electrical potential difference of corneal endothelium (TEPD) is linearly related to the fluid secretion which is characteristic of this epithelium [1; 2], and that the coupling between current and fluid movement [3; 4] is of electro-osmotic nature. Hence, determination of the TEPD would serve as a measure of the fluid movement. The oscillatory nature of the TEPD has been recognized recently [5]. In that paper, we use the Fourier transform to analyze the TEPD; the high-amplitude, low-frequency oscillations shown in the spectrum were linked to the operation of electrogenic sodium-bicarbonate cotransporters. However, no time localization of that activity could be resolved with the Fourier methodology utilized.

For that reason we now characterize the TEPD using wavelet analysis with the aim to localize in time the variations in TEPD. The main difference between these two methods is that the standard Fourier transform is only localized in frequency, whereas wavelets are localized in both frequency and time. We find that the high-amplitude oscillations of the TEPD take place throughout the experiment, with a mean period of 4.6 ± 0.4 s.

## MATERIALS AND METHODS

*Endothelial dissection and mounting.* Experiments were done using in vitro rabbit corneal endothelial preparations (posterior epithelium of the cornea). Techniques for the recording of electrical potential difference across it have been described previously [6; 7]. Corneas were obtained from New Zealand albino rabbits (~2 kg) using institutional procedures. Rabbits were euthanized by injecting a sodium pentobarbital solution into the marginal ear vein. The eyes were enucleated immediately. The cornea was deepithelialized and dissected using the method of Dikstein and Maurice [8], and was mounted in a jacketed Ussing chamber (T = 37 C). The endothelial and stromal surfaces were bathed with an air-bubbled HEPES-$HCO_3^-$ solution containing (in mmol/L): 104.4 NaCl, 26.2 $NaHCO_3$, 3.8 KCl, 1 $NaH_2PO_4$, 0.78 $MgSO_4$, 1.7 $CaCl_2$, 6.9 glucose, and 20 HEPES Na. The cornea was supported by a hemispherical stainless steel net, and subject to a hydrostatic pressure difference of 3 cm $H_2O$ applied to the aqueous (endothelial) side.

*Measurements of transendothelial electrical potential difference.* The TEPD was determined with a differential electrometer amplifier (Model 604; Keithley Instruments Inc., Cleveland, OH) connected to calomel electrodes and salt bridges. Small electrical asymmetries between them were nullified with an adjustable series battery. The TEPD (typically 0.5 – 1.5 mV) was amplified by the electrometer by $10^3$ and sent through a filter (HI/LO 1020F; Rockland Labs, Tappan, NY; low pass cutoff = 10 KHz). This processed signal was displayed on an oscilloscope, digitized at 500 Hz with a data acquisition system for electrophysiology recordings (Digidata 1440A, Axon Instruments,

Forest City, CA), and stored in a PC with the help of the program pClamp (Axon Instruments) . Each experiment was done in a different corneal preparation.

*Transendothelial electrical potential difference temporal records analysis.*

*Wavelet transform*

The wavelet transform (WT) is a time-scale (or time-frequency) transformation of continuous or discrete signals, and is more general than the windowed Fourier transform. It allows a multi-resolution analysis of the signal in the time-scale field, showing good resolution in time for high frequencies and good resolution in frequency for low frequencies [9; 10].

The normalized wavelet function is defined as:

$$\Psi_{a,b}(t) = \frac{1}{\sqrt{|a|}}\Psi\left(\frac{t-b}{a}\right) \qquad (1)$$

where $\Psi(t)$ is the mother wavelet function, and *a* and *b* are dilation and location parameters.

There are some requirements to consider a function as a wavelet. It must have finite energy; a mean value of zero; and for complex wavelets, the Fourier transform must be real, and must vanish for negative frequencies.

The continuous WT of a continuous signal *f(t)* is defined as

$$WT(a,b) = \int_{-\infty}^{\infty} f(t)\Psi^{*}{}_{a,b}(t)dt\ ,$$

where $\Psi^{*}{}_{a,b}$ is the complex conjugate of the function $\Psi_{a,b}$.

A discretization of the transform integrals can be performed to solve the continuous WT of a discretized signal *f(t)*, replacing the continuous integral by a discrete summation in time, with discrete evolution of parameters *a* and *b* where $t= n_t\Delta t$ and $b= n_b\Delta t$ .

$$WT(a,n_b)=\frac{1}{\sqrt{a}}\sum_{n_t=0}^{N-1} f(n_t\Delta t)\Psi^*_{a,b}\left(\frac{(n_t-n_b)}{a}\Delta t\right),\quad \text{with}\quad n_b=0\ldots N-1$$

Another form to show the wavelet transform is by means of the convolution in time between the measured signal and the mother wavelet.

$$WT(a,b)=f(t)\otimes\Psi^*_{a,b}(t)$$

Using the convolution theorem [11], the wavelet transform can be expressed as:

$$WT(a)=IFFT(\frac{1}{\sqrt{a}}FFT(f(n,\Delta t))FFT(\Psi(n,\Delta t,a)))$$

The benefit of the latest expression is the computational speed of resolution using FFT. The aim of these equations is to find the wavelet coefficients (or amplitudes) as a function of time and scale (which is related to frequency). The wavelet power spectrum is defined as the absolute-value squared of the wavelet coefficients.

*Analysis of the TEPD records.*

Wavelet analysis of TEPD was done using two routines written in Matlab. The first routine operated on the temporal record of the TEPD, and yielded the amplitudes of all frequencies up to 15 Hz. In this work, the Morlet function was selected as the mother wavelet. To concentrate on the analysis of the frequencies of most interest, a second routine was developed that considered only the frequencies between 0.5 and 2 Hz. It averaged the squared absolute values of the amplitudes (wavelet power) in the frequency domain, and gave as output a table of the times and the averaged wavelet power.

We subsequently took the temporal sequence of the wavelet power, and run the "Origin" pick peaks routine, with parameters of width 5, height 5, and minimum height 5%, thus determining the time and amplitude of the peaks for a given experiment. We then constructed the two histograms, (1) for the time differences between the peaks (in s), and (2) for their wavelet powers (in $Volt^2$ Hz). We obtained fits for them using Origin lognormal distribution. Results are expressed as mean ± SE.

## RESULTS

*Temporal records of TEPD and wavelet analysis*

We obtained temporal records of the electrical potential difference between the two experimental hemichambers filled with saline, with (n=4) and without (n=3) corneal preparations mounted in between them. As was reported earlier [5], a characteristic behavior of temporal records of the TEPD was observed only with preparations present (Fig. 1 top panel). By comparison, controls with only solution in the chambers were remarkably flat. Subsequently, we performed a wavelet analysis of the temporal record of the TEPD and we obtained a 3D graph (Fig. 1) depicting the frequency (in the ordinate); the time (on the abscissa); and in the z axis (towards the observer), the amplitude color-coded with the red being high amplitude (blue = background; Fig. 1 bottom panel).

Inspection of these figures revealed most often a periodicity; as Fig. 1 shows, low-frequency amplitudes were present in packets or groups occurring every ~4-5 s. Based on this qualitative perception, we devised the analytical procedure that follows.

*Analysis of the wavelet power of the TEPD*

We limit our analysis to the low frequencies of most interest (0.5-2 Hz), and we determine the time and the wavelet power peaks for each experiment. For convenience, only the first 195 s of an experiment are shown in figure 2. In the time domain, the peaks correspond to periods of ~ 5 seconds (Fig. 2). In contrast, only a flat baseline with negligible noise was obtained in control experiments generated by the solution-filled chamber, which rules out possible artifacts from electrodes, electrometer or unspecified sources (data not shown).

To quantify these results we obtained the histograms for the time differences between peaks ($\Delta t$), and for their wavelet powers. Figure 3 shows the histogram for the $\Delta t$'s obtained from the data shown in figure 2. A fit to the lognormal distribution yields for that experiment a time interval of 4.8 ± 0.3 s (0.21 Hz). The wavelet power peak amplitudes of figure 2 were in turn grouped into a histogram and fitted again to the lognormal distribution (Fig. 4), which yielded an amplitude of 2.1 ± 0.1 $mV^2$ Hz. A summary of the values for $\Delta t$ and wavelet power peak amplitudes for all four experiments is given in Table 1.

## DISCUSSION

The oscillatory nature of the TEPD recognized recently was achieved with Fourier analysis. What that methodology could not resolve was when in time did the oscillations appear during the experiment. This limitation is now circumvented by the use of wavelet analysis. The most important conclusion is that the high-amplitude oscillations of the TEPD occur all throughout the experiment, without significant calm or rest periods, repeating themselves with a period of (Table 1) 4.6 ± 0.4 s (0.22 Hz). The wavelet power at the peak of such oscillations is 1.5 ± 0.1 $mV^2$ Hz (Table 1) in the average, and is remarkably narrow in its distribution.

Low frequency activity in the TEPD was attributed to the functioning of sodium-bicarbonate cotransporters in a recent publication of ours [5] which analyzed the power spectrum of the TEPD. The current observations with wavelet analysis confirm and extend the prior findings; the low-frequency oscillations of the TEPD, attributable to sodium-bicarbonate cotransport, occur in the fashion described above.

We speculate that cotransporters are recruited across the cells by some unspecified signal with the periodicity noted. Further implications are that neither the number of cotransporters nor the number of charges transported varies much in each cycle.

*The electrogenic sodium-bicarbonate transporter*.

There is electrophysiological evidence for sodium-bicarbonate cotransport in the corneal endothelium. The existence of a symport that transports sodium and bicarbonate out of the cell at the apical membrane in this preparation is strongly supported by the electrophysiological evidence of the Wiederholt laboratory [12; 13; 14; 15; 16]. Immunocytochemical localization found cotransporters in both the basolateral and apical membranes [17].

As a result of the operation of the sodium-bicarbonate cotransporters in tandem, bicarbonate enters the cell basolaterally, and exits apically [17]. This generates an electrical potential difference (aqueous negative) that would in turn result in a flow of positive charges ($Na^+$ ions) through the junction. It is such flow that, coupled to water by electro-osmosis [3; 4], finally results in fluid transport through this preparation.

**Acknowledgment.**

This work was supported by CONICET PIP # 1688 (to JF). NM was a doctoral fellow (CONICET).

Table 1.

| Experiment | Wavelet power ($mV^2$ Hz) | Δt (sec) |
|---|---|---|
| 1 | 2.13 ± 0.07 | 4.81 ± 0.27 |
| 2 | 1.48 ± 0.08 | 4.38 ± 0.26 |
| 3 | 0.95 ± 0.03 | 4.69 ± 0.09 |
| 4 | 1.56 ± 0.05 | 4.54 ± 0.18 |
| Mean ± SE | 1.53 ± 0.12 | 4.59 ± 0.43 |

Summary of wavelet power peak amplitudes, and Δt's of all experiments.

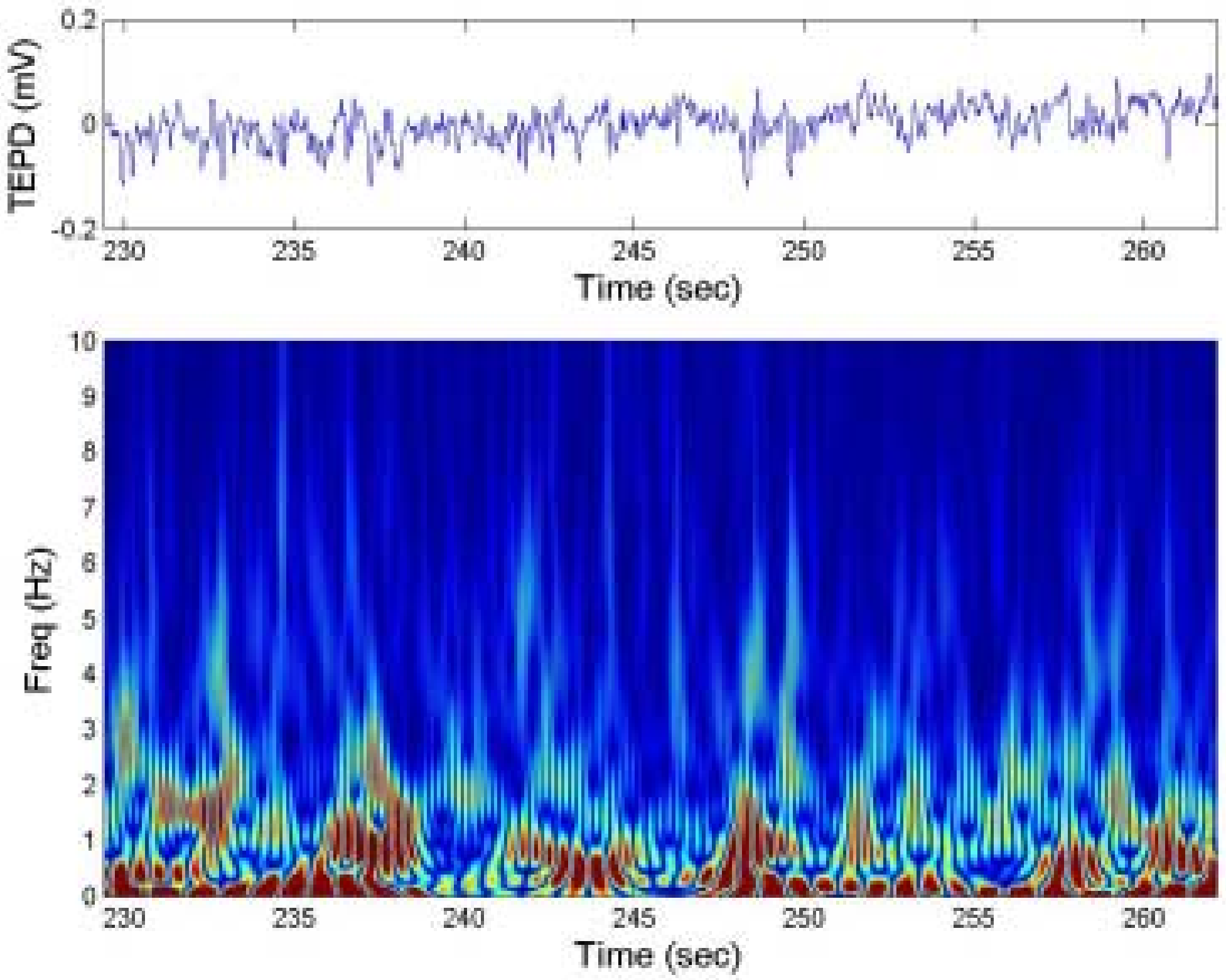

Figure 1. Wavelet analysis of the temporal record of TEPD (representative experiment).
Top panel: transendothelial electrical potential difference (in mV) as a function of time. As time increases, the TEPD goes down gradually as the preparation decays.
Bottom panel: 3D graph depicting: in the ordinate, the frequency; on the abscissa, the time; in the z axis (towards the observer), the amplitude color-coded with the red being high amplitude and the blue denoting the background.
The bottom panel was obtained from the data in the top panel, treated with the wavelet-decomposition routine run in Matlab.

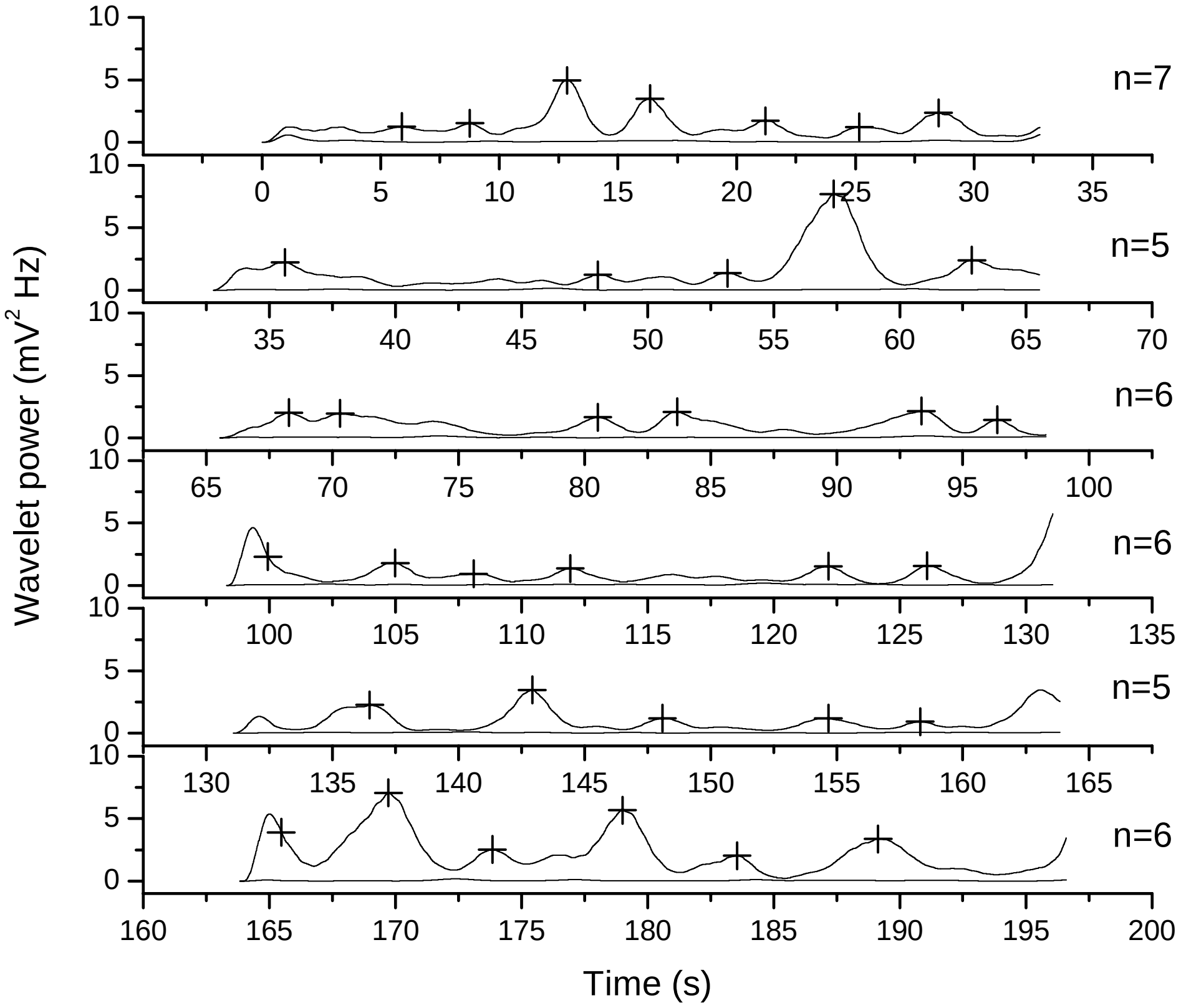

Figure 2. Representative experiment of wavelet power. Of all the data (between 0 and 15 Hz), the Matlab routine selected only the region of highest wavelet power (between 0.5 and 2 Hz), squared them, and averaged them. The resulting wavelet power is plotted against time for the first 197 s. The crosses denote the times at which peaks were found by Origin®; n is the number of peaks in each row.

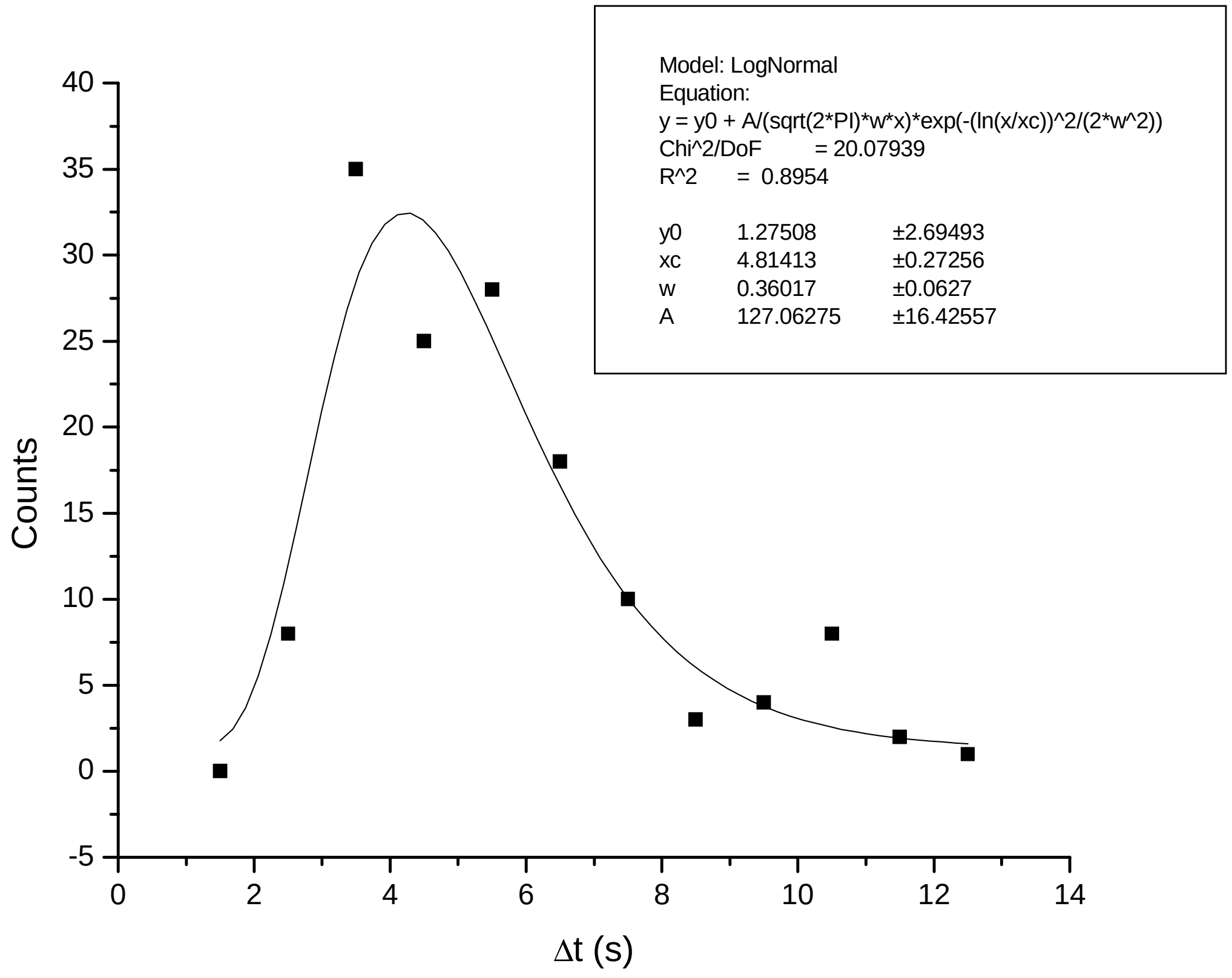

Figure 3. The time intervals between peaks for experiment 1 in Fig. 2 ($\Delta t$'s) are grouped in a histogram. A fit to the lognormal distribution (done in Origin®) yields an average time interval of $4.8 \pm 0.3$ s.

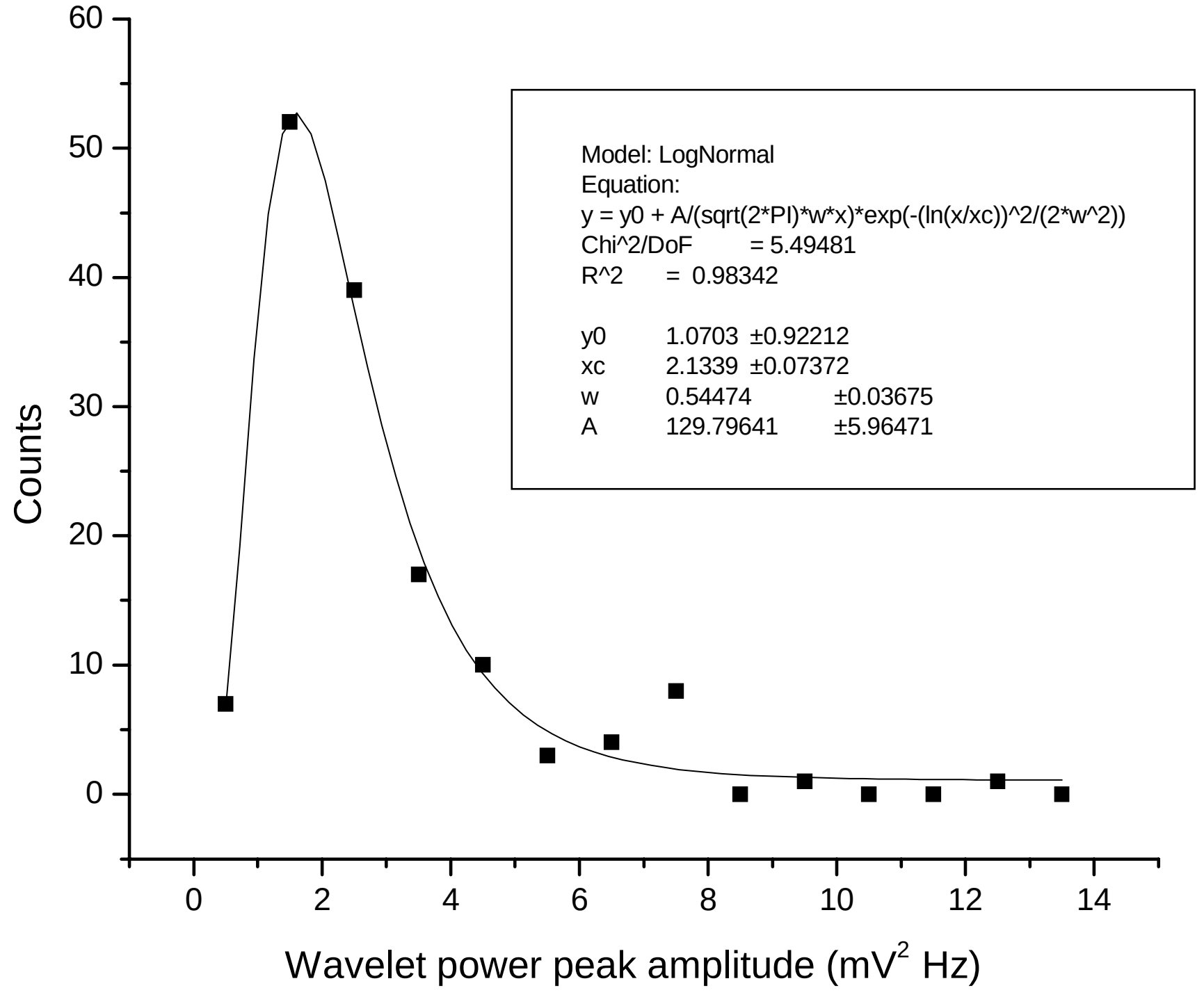

Figure 4. The wavelet power peak amplitudes of Fig. 2 are grouped into a histogram and fitted to the lognormal distribution by Origin ®. The average value is $2.1 \pm 0.1$ $mV^2$ Hz ; note the relative narrowness of the distribution.

REFRENCES